\documentclass[final]{svjour3}
\usepackage[pdftex]{graphicx}
\usepackage{rotating}
\usepackage{amssymb}
\usepackage{amsmath}
\usepackage{mathptmx}
\usepackage[numbers]{natbib}
\makeatletter
\journalname{Journal of Low Temperature Physics}

\makeatletter
\newcommand\subparagraph{%
  \@startsection{subparagraph}{5}
  {\parindent}
  {3.25ex \@plus 1ex \@minus .2ex}
  {-1em}
  {\normalfont\normalsize\bfseries}}
\makeatother
\usepackage{titlesec}
\let\subparagraph\relax 
\titlespacing\section{0pt}{3.5ex plus 1ex minus .2ex}{1.5ex plus .2ex}
\titlespacing*{\subsection}   {0pt}{3.25ex plus 1ex minus .2ex}{1.5ex plus .2ex}
\titlespacing*{\subsubsection}{0pt}{3.25ex plus 1ex minus .2ex}{1.5ex plus .2ex}	
\newcommand*{\citen}[1]{%
  \begingroup
    \romannumeral-`\x 
    \setcitestyle{numbers}%
    \cite{#1}%
  \endgroup   
}

\bibpunct{}{}{,}{s}{}{,}

\begin{document}

\newcommand{\hdblarrow}{H\makebox[0.9ex][l]{$\downdownarrows$}-}
\title{SLAC Microresonator Radio Frequency (SMuRF) Electronics for Read Out of Frequency-Division-Multiplexed Cryogenic Sensors}

\author{S.~A.~Kernasovskiy$^a$ \and 
S.~E.~Kuenstner$^a$ \and 
E.~Karpel$^a$ \and 
Z.~Ahmed$^{b,\dagger}$ \and 
D.~D.~Van~Winkle$^b$ \and 
S.~Smith$^b$ \and 
J.~Dusatko$^b$ \and 
J.~C.~Frisch$^b$ \and 
S.~Chaudhuri$^a$ \and 
H.~M.~Cho$^b$ \and 
B.~J.~Dober$^c$ \and 
S.~W.~Henderson$^b$ \and 
G.~C.~Hilton$^c$ \and 
J.~Hubmayr$^c$ \and 
K.~D.~Irwin$^{a,b}$ \and 
C.~L.~Kuo$^{a,b}$ \and 
D.~Li$^b$ \and 
J.~A.~B.~Mates$^c$ \and 
M.~Nasr$^b$ \and 
S.~Tantawi$^b$ \and 
J.~Ullom$^c$ \and 
L.~Vale$^c$ \and 
B.~Young$^d$}

\institute{
$^a$Department of Physics, Stanford University,\\ Stanford, CA 94305, USA\\
$^b$SLAC National Accelerator Laboratory,\\ Menlo Park, CA 94025, USA\\
$^c$National Institute for Standards and Technology,\\ Boulder, CO 80305, USA\\
$^d$Santa Clara University,\\
Santa Clara, CA 95053, USA\\
$^{\dagger}$\email{zeesh@slac.stanford.edu}\\}

\maketitle

\begin{abstract}
Large arrays of cryogenic sensors for various imaging applications ranging across x-ray, gamma-ray, Cosmic Microwave Background (CMB), mm/sub-mm, as well as particle detection increasingly rely on superconducting microresonators for high multiplexing factors. These microresonators take the form of microwave SQUIDs that couple to Transition-Edge Sensors (TES) or Microwave Kinetic Inductance Detectors (MKIDs).  In principle, such arrays can be read out with vastly scalable software-defined radio using suitable FPGAs, ADCs and DACs. In this work, we share plans and show initial results for SLAC Microresonator Radio Frequency (SMuRF) electronics, a next-generation control and readout system for superconducting microresonators. SMuRF electronics are unique in their implementation of specialized algorithms for closed-loop tone tracking, which consists of fast feedback and feedforward to each resonator's excitation parameters based on transmission measurements. Closed-loop tone tracking enables improved system linearity, a significant increase in sensor count per readout line, and the possibility of overcoupled resonator designs for enhanced dynamic range. Low-bandwidth prototype electronics were used to demonstrate closed-loop tone tracking on twelve 300-kHz-wide microwave SQUID resonators, spaced at $\sim$6\,MHz with center frequencies $\sim$5--6\,GHz. We achieve multi-kHz tracking bandwidth and demonstrate that the noise floor of the electronics is subdominant to the noise intrinsic in the multiplexer.  


\end{abstract}

\section{Introduction}

The sensitivity of modern CMB bolometric sensors used in ground-based and space-based cameras is photon-noise-limited.  Thus a camera's  
sensitivity scales approximately as the square root  of the number of sensors or of the integration time.  To keep integration times reasonable, 
sensor counts deployed on CMB cameras have been dramatically 
increasing, from only a handful of sensors two decades ago to a projected nearly 500,000 
sensors in CMB-S4 [\citen{abitbol2017s4Tech}].  
X-ray imaging spectrometers based on Transition Edge Sensor (TES) microcalorimeters have also increased sensor count in order to increase the photon collection efficiency of the TES array without degrading energy resolution [\citen{ullom2015gammaArray}]. 

The sensors are operated at cryogenic temperatures, 
thus heat load from room-temperature wires must be reduced by 
multiplexing sensor signals at the cold stages.  Common forms of 
multiplexing for TES bolometers, such as time-division multiplexing using Superconducting Quantum Interference devices (SQUIDs) and amplitude-modulated frequency-division multiplexing, have reached multiplexing factors of 
$\sim$65 in deployed CMB cameras [\citen{dobbs2009TDM,benson2014spt,Henderson2015}].  

Superconducting microresonators with resonator frequencies of order 0.1--10\,GHz can be excited and probed in the thousands, by coupling to a single RF probe line  [\citen{zmuidzinas2012,irwin2009shannon}]. 
The resonators can double as photon sensors in the case of Microwave Kinetic Inductance 
Detectors (MKIDs), where Cooper-pair breaking by photons influences resonance amplitude and frequency in a measurable way [\citen{Day2003}]. Alternatively, the resonators can couple to TESs via microwave SQUIDs, which transduce the current changes in the TESs into frequency shifts [\citen{mates2008}]. In both schemes, the resonators can be excited and 
read out in large numbers with software-defined radio (SDR) using 
suitable FPGAs, ADCs and DACs. Several groups are using SDR hardware to develop readout systems for microresonators [\citen{werthimer2011casper}].  

This paper will describe a next-generation readout system called the
 SLAC Microresonator Radio Frequency (SMuRF) electronics specifically designed to read out more than 4,000 channels
in 4\,GHz of bandwidth (Section~\ref{sec:elecdesc}).  SMuRF is unique in that it implements 
fast tracking of resonances, which reduces the power to the 
follow-on amplifier, thereby increasing dynamic range or enabling a large multiplexing factor (Section~\ref{sec:tone}). We show that the readout noise of a prototype for SMuRF is subdominant to 
device noise (Section~\ref{sec:noise}).

\section{SMuRF electronics description}
\label{sec:elecdesc}

\begin{figure}[htbp]
\includegraphics[width=\textwidth]{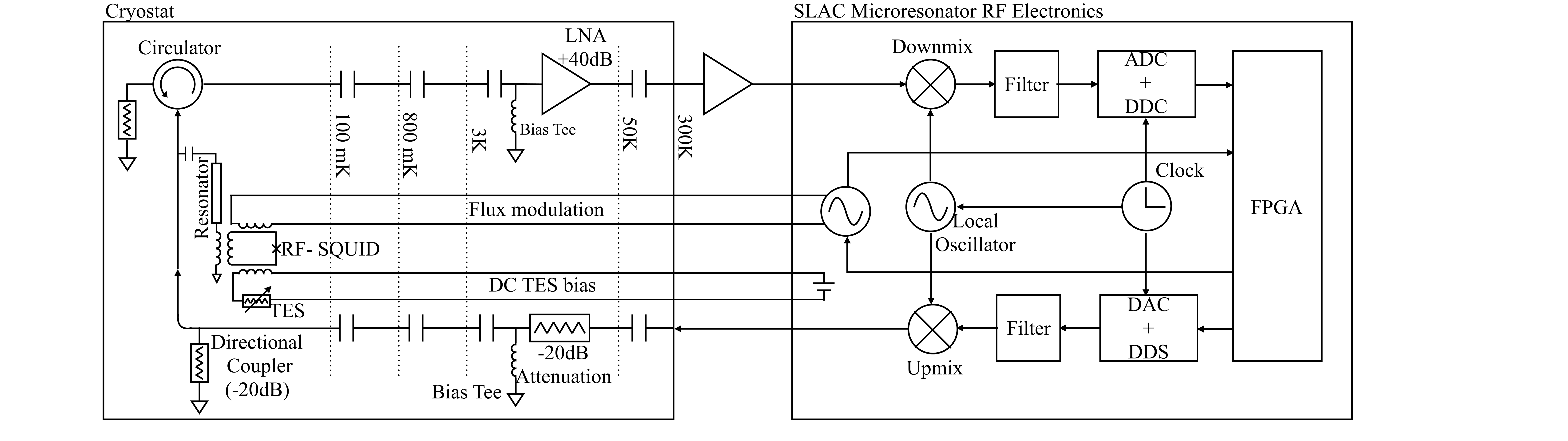}
\caption{Schematic of a typical readout setup with 
microwave SQUIDs coupled to TESs. On the right in the electronics, tones are continuously generated and updated in the FPGA at baseband and up/down-mixed to feed into or read back from the cryostat. The cryogenic resonances on the left, in this case, superconducting microresonators coupled to RF-SQUIDS, are probed by the tones from the electronics. TES signals shift the fundamental frequency of the resonator. By modulating the SQUIDs, the TES signal becomes phase-modulated, which the electronics then read back, demodulate and use to update the tones.}
\label{fig:schematic}
\end{figure}

A schematic of the SMuRF readout system with a typical 
setup of microwave SQUIDs coupled to TESs is shown in 
Fig.~\ref{fig:schematic}. At the heart of the system is an FPGA running firmware for tone generation, tracking, read out and signal demodulation.
The system uses a digital I/Q scheme to avoid analog electronic 
calibration errors.
Excitation tones are generated at baseband by the DAC using Direct Digital 
Synthesis (DDS).  Those tones are upmixed to the RF band, sent into the cryostat to probe the 
resonances, and the transmitted signals are then downmixed to baseband.  The ADC uses Digital Down Conversion (DDC) to measure the 
phase shift of the tones after they interact with the 
microresonators.  An updated frequency is then synthesized in the FPGA to drive 
each resonator at its new resonant frequency. 

\vspace{-\baselineskip}

\subsection{Physical architecture}
\label{sec:smurf_arch}
\begin{figure}[htbp]
\centering
\includegraphics[width=\textwidth]{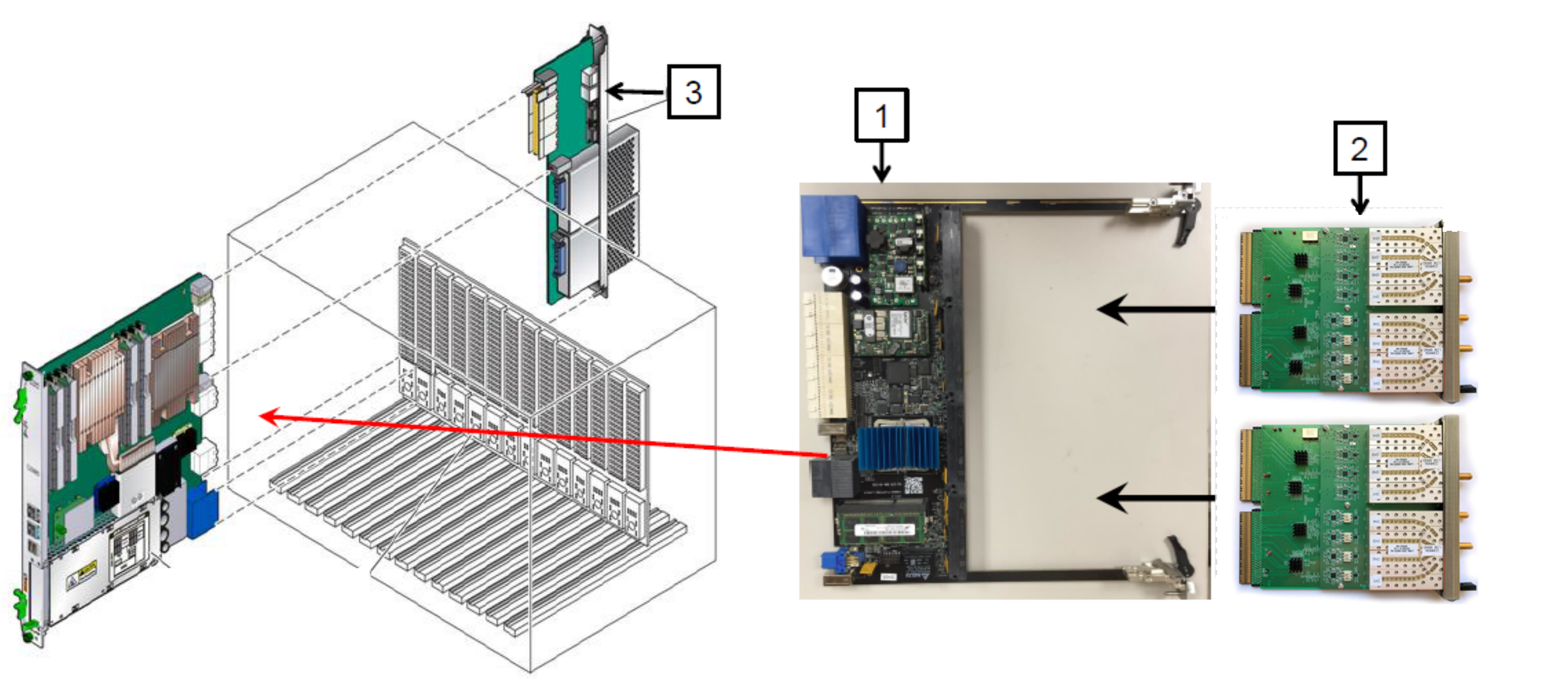}
\caption{Illustration of the layout of the readout card.  The FPGA 
carrier card is labeled 1.  The AMC mezzanine cards, labeled 2, 
each contain four DAC/ADC pairs.  An RF daughter card sits on top of the 
AMC card to define the microresonator bands, allowing for flexibility 
in choice of band.  The rear-transition module, labeled 3, contains 
all the low-frequency functions, including TES bias and flux 
bias for SQUIDs.}
\label{fig:eleclayout}
\end{figure}

The SMuRF system is built on the Advanced Telecommunication 
Computing Architecture (ATCA) specification. The physical layout 
is shown in Fig.~\ref{fig:eleclayout}.  
The base unit of the SMuRF system is a single slot in an ATCA 
crate, which houses a carrier card.
A carrier card contains a Xilinx Kintex Ultrascale FPGA, DDR memory, connections to a 40\,Gbps backplane, and connections to two smaller boards called Advanced Mezzanine Cards (AMCs). The carrier card used for SMuRF was developed at SLAC as part of a 
``Common FPGA platform'' [\citen{herbst2014design}].

The FPGA is connected to two AMCs with 16 
bi-directional 12.5~Gbit JESD lanes. The AMCs each contain 4 DAC/ADC pairs
and local oscillators (LO).  Each AMC is connected to a radio 
frequency (RF) daughter card, which carries bandpass filters and 
mixers to upconvert the tones to the microresonator band. This is chosen 
to be on a daughter card  
to allow flexibility in choosing the microresonator band.

Finally, a rear transition module (RTM) connected to the carrier 
card via the backplane handles low-frequency functions, 
including 32 independent TES bias channels and flux biasing for microwave SQUIDs.  It also 
provides stable biases for the low-noise amplifiers in the signal chain. 
The flux bias can be differentially 
driven by a slow (20-bit, 1\,Ms/s) DAC or a fast 
(16~bit, 50~Ms/s) DAC, for modulating signals from 
CMB bolometers and x-ray calorimeters, respectively. 
For experiments to use SMuRF electronics, a simple, customized interface card with passive analog elements connects wiring between the cryostat and the SMuRF crates. These interface boards will enable setting  bias ranges, connectors for the low-frequency wiring, grounding schemes, and low-pass filtering as desired. 

\vspace{-\baselineskip}

\subsection{Comb generation}

A comb of tones corresponding to resonance frequencies in a 500\,MHz block is generated on the FPGA at baseband 
(750~MHz-1.25~GHz). 
For each resonator, the FPGA synthesizes 
a high-amplitude drive tone centered directly on the resonance, and optionally two lower-amplitude 
sidebands detuned by a half-bandwidth for calibration as necessary. 
The comb of output tones is played out as a pair of 625\,MHz 
complex timestreams by a 16~bit, 2.5~Gs/s DAC by DDS. Local oscillators (LO) are used to mix the baseband comb up 
to the microresonator RF frequencies.  
On-board frequency multiplexers combine 8 such 500-MHz blocks into a single 4\,GHz band, 
which is routed to the cryostat on a 
single coaxial cable. Upon return to the electronics, the 4\,GHz band is channelized into 
8 $\times$ 500\,MHz bands using bandpass filters. 
The tones in each band are downmixed to baseband using the same LOs, and 
digitized and channelized by a 14~bit, 2.5~Gs/s ADC using DDC. 

A different choice of LO frequencies and bandpass filters would allow the output tones to 
access a different RF frequency if the microresonators were,
for instance, in the 1--2~GHz band rather than the 4--8~GHz band. Bypassing the mixers would allow direct sampling and excitation of 0.5--1~GHz band resonators.

\subsection{Prototype system}

A prototype system with a standard SLAC FPGA carrier board, but with ADCs and DACs to manipulate only 150\,MHz of bandwidth was assembled to test the tracking algorithms and  
noise properties. 
Low-frequency signals, various RF reference signals and filtering and mixing functions in this prototype 
were performed by benchtop electronics, instead of the compact form factor being developed for the full SMuRF electronics described in Section \ref{sec:smurf_arch}.
Measurements were performed on two NIST 
$\mu$mux16b microwave SQUID chips containing 64 resonances with 
bandwidths of 300~kHz between 5.0 
and 5.5~GHz [\citen{dober2017}]. 

\section{Tone tracking}
\label{sec:tone}

A unique feature of the SMuRF electronics is the ability to track resonator dips and adjust excitation tones at high bandwidth.  This reduces the power transmitted to
the follow-on low-noise amplifier since the tone is always located at the frequency 
of lowest transmission.  This in turn increases the amplifier's dynamic range, reduces the linearity requirements 
of the system, and allows more resonances to be read out 
with the same amplification chain.
With the prototype setup we were able to demonstrate fast tone tracking on twelve
channels simultaneously. 

\begin{figure}[htbp]
\centering
\includegraphics[width=0.49 \textwidth]{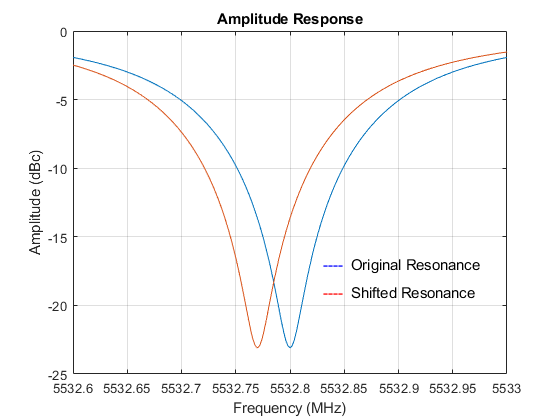}
\includegraphics[width=0.49 \textwidth]{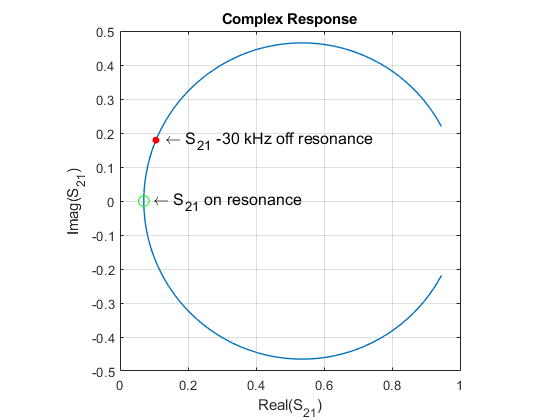}
\caption{Simulated response of a resonator-loaded transmission line near a resonance at 5532.8 MHz.
\textit{Left}: Amplitude response for resonator at original frequency (blue) and after a frequency shift (red).
\textit{Right}: Complex response (after phase rotation) so that frequency difference between the probe frequency and the resonance minimum is proportional to the imaginary component of the complex frequency response S21. The latter can therefore be used to feedback and correct the probe frequency.       
}
\label{fig:IQrot}
\end{figure}

The received signal in a narrow band around each resonance line is demodulated to a complex
amplitude. 
This amplitude is phase-rotated so the quadrature component is proportional to the
frequency deviation from the center of the resonance line. 
The frequency error estimate, updated at 1.3~MHz for each resonance, is input to a feedback loop filter which updates the digitally-generated frequency of each tone used to drive the multiplexer array. 
Two low-amplitude ($\sim$20~dB down) sidebands are generated around each line-tracking tone in order to calibrate the phase rotation and scale needed to translate complex amplitude to frequency error. 
The complex amplitudes of these sidebands, normalized to their generated amplitudes, are averaged to
estimate a reference approximating the response of the transmission in the absence of a resonator
notch. 
Initial estimates of resonator parameters are taken either from a network analyzer or
amplitude scan with the readout electronics.  
The process of updating the tones is illustrated in Fig.~\ref{fig:IQrot}.

\vspace{-\baselineskip}

\subsection{Tone tracking bandwidth}

The bandwidth of the tone tracking algorithm was 
measured for the prototype electronics by abruptly 
changing the frequency of the center tone and recording the response.  
The bandwidth was measured to be 
about 20~kHz.  This was limited by the algorithmic latency 
of the prototype system and can be improved.

While 20~kHz tone tracking is adequate for most CMB applications, 
the x-ray applications need a bandwidth of closer to 1~MHz.  A feed-forward
algorithm is being developed in which the resonator movements
are predicted for flux bias at MHz speed and the tones are updated 
at that rate,
while the feedback occurs at a slower rate.


\subsection{Reduction of power}
\begin{figure}[htbp]
\centering
\includegraphics[width=0.45 \textwidth]{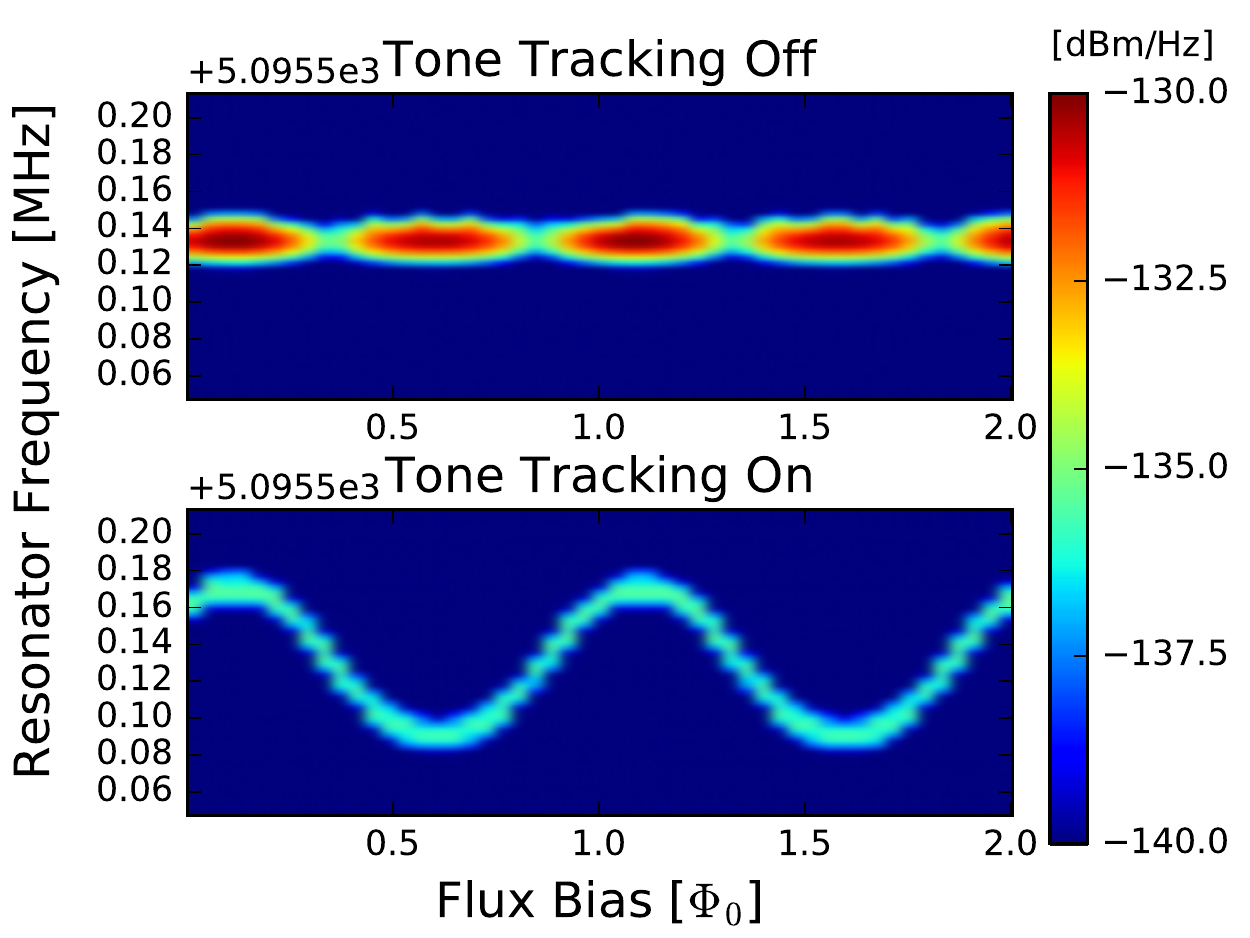}
\includegraphics[width=0.45 \textwidth]{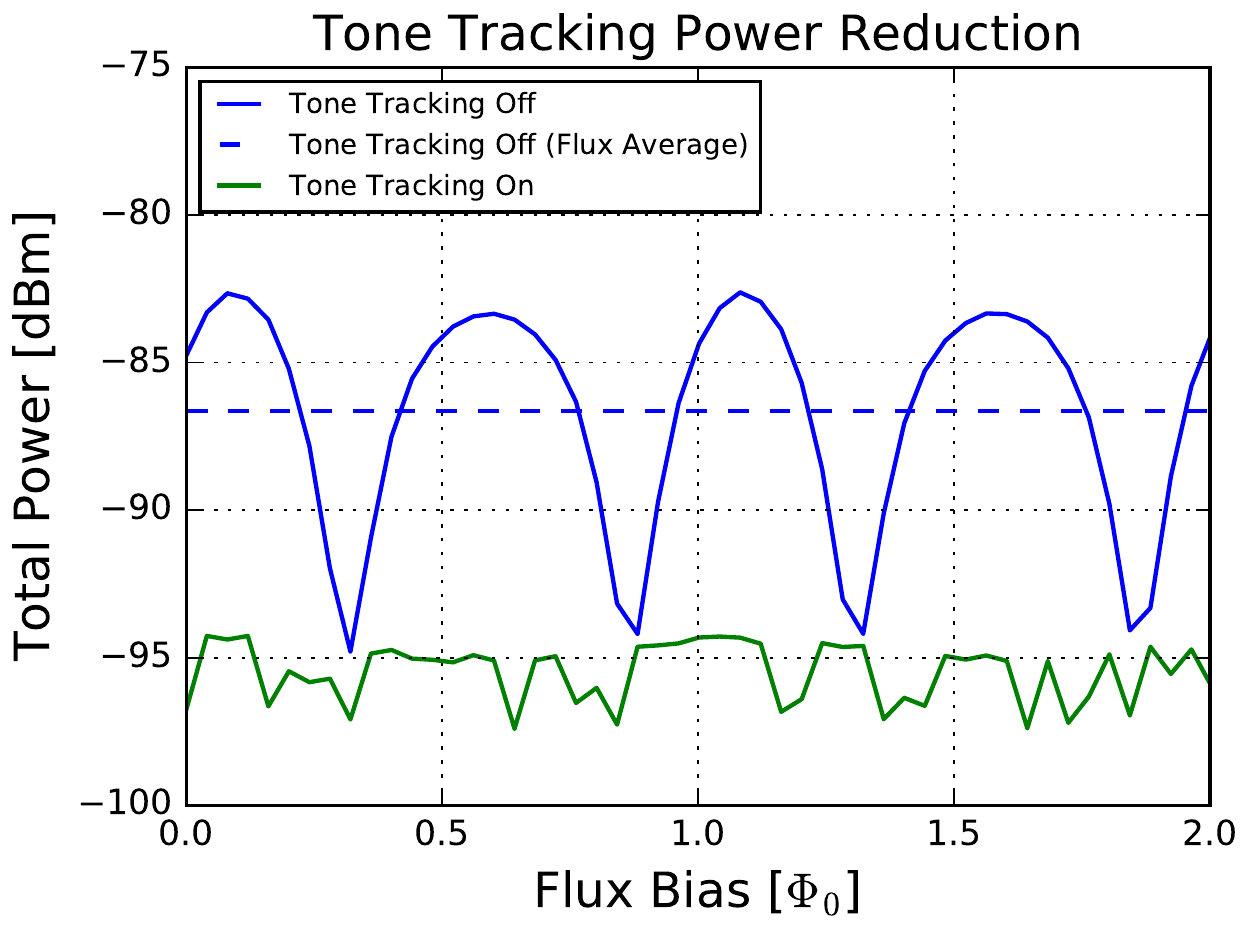}
\caption{\textit{Left} the spectrum for a single line read out in open and 
closed-loop mode. In closed-loop mode, the frequency of the tone shifts to 
follow the resonance as the flux signal changes the resonance frequency. 
In order to show the power differences more clearly, the two sidebands are 
not shown. The quantization visible in the closed-loop plot is due to the 
small number of input fluxes tested, and is not due to discretization of the 
tracking frequencies. \textit{Right} The total RF power for a single line as a function 
of input flux, referred to the input of the HEMT amplifier. The total power 
is obtained by integrating along the vertical axis in the left plot. The 
dashed blue line shows the flux-averaged single resonance power.}
\label{fig:powerreduc}
\end{figure}

Tone tracking enables most of the 
drive power to be reflected away from the input of the HEMT amplifier. 
To demonstrate the improved linearity of the system in 
closed-loop mode, power at the HEMT input was measured while 
a sawtooth flux bias of amplitude $2\Phi_0$ was applied to the common flux bias line. 
This modulated all resonance frequencies simultaneously. The drive tones follow resonators as they shift, and the power 
transmitted to the HEMT input remains constant as a function of 
flux bias. In open-loop mode, the drive tones are held fixed at the 
flux-averaged frequency centers of the resonators, and the resonators shift past 
the drive tones, as shown in Fig.~\ref{fig:powerreduc}. Averaged 
over a flux bias period, the power per resonance is 
$\sim$ 9\,dB lower in closed-loop mode for this set of 
resonators, as expected. 

\vspace{-\baselineskip}

\section{Noise performance}
\label{sec:noise}

\begin{figure}[htbp]
\centering
\includegraphics[width=0.45 \textwidth]{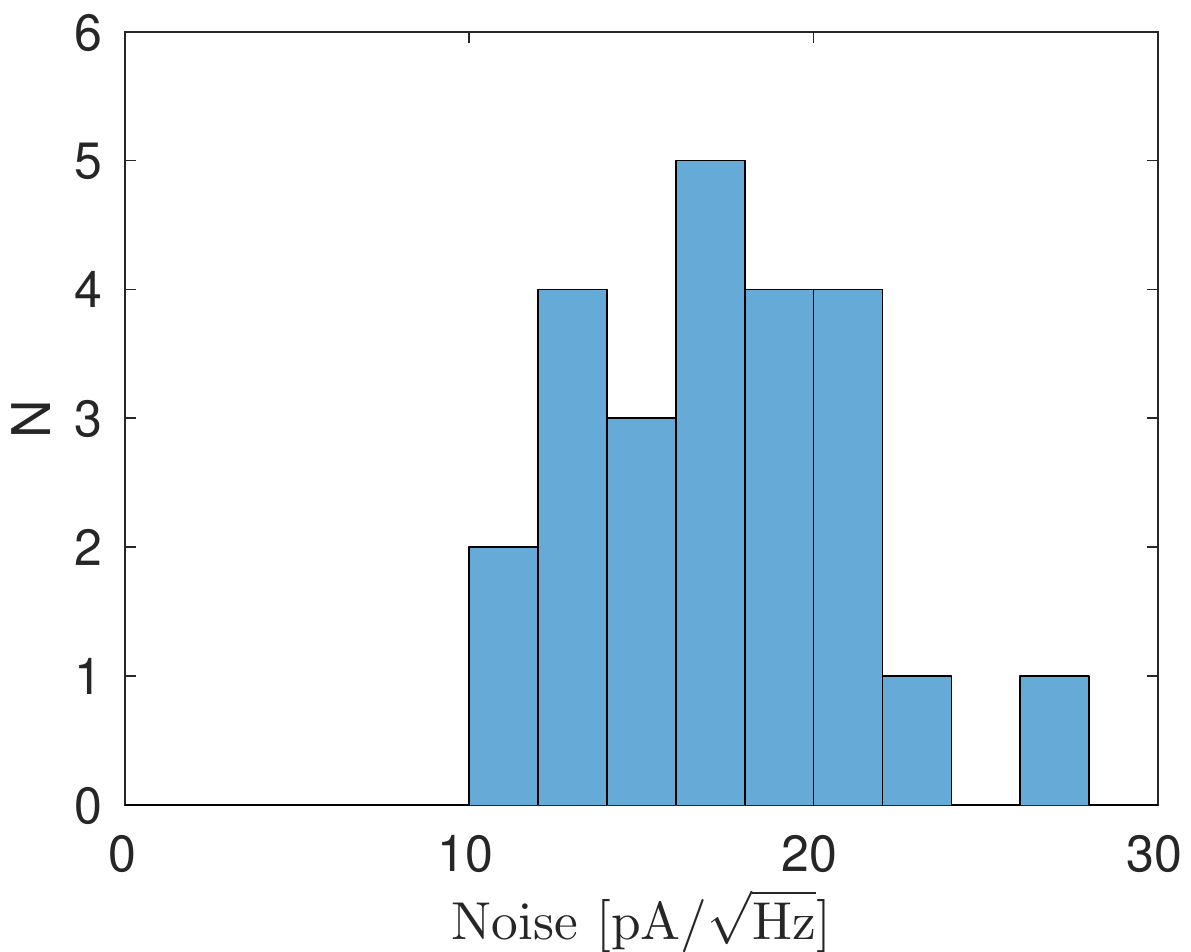}
\caption{ Histogram of the noise of a $\mu$mux16b microwave SQUID chip measured with a DC flux bias.  
This in agreement with independent measurements of the SQUID noise for these same chips.
}
\label{fig:noisehist}
\end{figure}

Because the prototype was not optimized for phase noise, we did not evaluate it directly, but this is an important measurement for the full 4\,GHz bandwidth SMuRF system. 
We did, however, characterize the system noise of the prototype running closed-loop tone tracking on $\mu$mux16b microwave SQUID multiplexers. 
The noise was measured with a DC flux bias and 
referred to the input current using 
the slope of the SQUID response curve and the relative 
mutual inductance of the input line 
compared to the flux bias line.  This is more 
sensitive to input current than when 
averaged over a period of AC flux bias.
The measured noise, seen in Fig.~\ref{fig:noisehist}, is between 10--20~pA/$\sqrt{\mathrm{Hz}}$ 
as observed in measurements of the same type of chips 
in other systems [\citen{dober2017}].



\vspace{-\baselineskip}

\section{Conclusion}

Using an early prototype of SMuRF electronics we have demonstrated tone tracking 
over wide RF bandwidth suitable for CMB camera readout.  
Tone tracking is unique to the SMuRF system and 
significantly reduces the linearity requirements 
on the cold and warm readout, potentially enabling 
larger multiplexing factors than achievable 
with otherwise comparable systems.
The first full SMuRF system is being assembled 
now and is being designed to read out more than 4000
channels in 4~GHz of bandwidth.

\begin{acknowledgements}
This work was supported by the Department of Energy Office of Science Detector R\&D funds.
\end{acknowledgements}

\pagebreak

\bibliographystyle{ieeetr}
\bibliography{citations}

\end{document}